\documentclass[3p,twocolumn,times]{elsarticle}

\usepackage{amsmath}
\usepackage{amssymb}
\usepackage{ragged2e}
\usepackage{tikz}
\usepackage{caption}
\usepackage{subcaption}
\usepackage{booktabs}
\usepackage[hyphens]{url}
\usepackage{breakurl}
\usepackage[breaklinks]{hyperref}

\usepackage{pifont,xspace}
\newcommand{\cmark}{\ding{51}\xspace}%
\newcommand{\xmark}{\ding{55}\xspace}%

\newcounter{x}
\newtheorem{axiom}{Axiom A$\!\!$}

\usetikzlibrary{arrows,patterns,automata,backgrounds,decorations,fit,petri,positioning,petri,shapes,calc}
\tikzset{place/.append style={circle,draw=black,thick,inner sep=0pt,minimum size=3mm,label position=below}}
\tikzset{transition/.append style={rectangle,draw=black,thick,inner sep=1pt,minimum size=4mm}}
\tikzset{every edge/.append style={-{>[sep=0pt]}, thin}}
\tikzset{pre/.append style={<-,shorten <=0pt,shorten >=0pt}}
\tikzset{post/.append style={->,shorten >=0pt,shorten <=0pt}}
\newtheorem{definition}{Definition}

\newcommand{\fleche}{\longrightarrow}
\newcommand{\flsup}[1]{\stackrel{#1}{\fleche}}
\newcommand{\step}[1]{\flsup{#1}}           %   -e->
\newcommand{\Lan}                 {\mathfrak{L}}

\newcommand{\APN}{\mathit{APN}}

\newcommand{\MF}{\mathit{MF}}
\newlength{\hatchspread}
\newlength{\hatchthickness}
\newlength{\hatchshift}
\newcommand{\hatchcolor}{}
\tikzset{hatchspread/.code={\setlength{\hatchspread}{#1}},
	hatchthickness/.code={\setlength{\hatchthickness}{#1}},
	hatchshift/.code={\setlength{\hatchshift}{#1}},% must be >= 0
	hatchcolor/.code={\renewcommand{\hatchcolor}{#1}}}
\tikzset{hatchspread=3pt,
	hatchthickness=0.4pt,
	hatchshift=0pt,% must be >= 0
	hatchcolor=black}
\pgfdeclarepatternformonly[\hatchspread,\hatchthickness,\hatchshift,\hatchcolor]% variables
{custom north west lines}% name
{\pgfqpoint{\dimexpr-2\hatchthickness}{\dimexpr-2\hatchthickness}}% lower left corner
{\pgfqpoint{\dimexpr\hatchspread+2\hatchthickness}{\dimexpr\hatchspread+2\hatchthickness}}% upper right corner
{\pgfqpoint{\dimexpr\hatchspread}{\dimexpr\hatchspread}}% tile size
{% shape description
	\pgfsetlinewidth{\hatchthickness}
	\pgfpathmoveto{\pgfqpoint{0pt}{\dimexpr\hatchspread+\hatchshift}}
	\pgfpathlineto{\pgfqpoint{\dimexpr\hatchspread+0.15pt+\hatchshift}{-0.15pt}}
	\ifdim \hatchshift > 0pt
	\pgfpathmoveto{\pgfqpoint{0pt}{\hatchshift}}
	\pgfpathlineto{\pgfqpoint{\dimexpr0.15pt+\hatchshift}{-0.15pt}}
	\fi
	\pgfsetstrokecolor{\hatchcolor}
	\pgfusepath{stroke}
}

\journal{}

\begin{document}

\begin{frontmatter}



\title{The Imprecisions of Precision Measures in Process Mining}

\author{Niek Tax\corref{cor1}}
\cortext[cor1]{Corresponding author}
\ead{n.tax@tue.nl}
\author{Xixi Lu\corref{}}
\ead{x.lu@tue.nl}
\author{Natalia Sidorova\corref{}}
\ead{n.sidorova@tue.nl}
\author{Dirk Fahland\corref{}}
\ead{d.fahland@tue.nl}
\author{Wil M.P. van der Aalst\corref{}}
\ead{w.m.p.v.d.aalst@tue.nl}

\address{Eindhoven University of Technology, P.O. Box 513, Eindhoven, The Netherlands}

\begin{abstract}
In process mining, \emph{precision measures} are used to \emph{quantify} how much a process model overapproximates the behavior seen in an event log. Although several measures have been proposed throughout the years, no research has been done to validate whether these measures achieve the intended aim of quantifying over-approximation in a consistent way for all models and logs. This paper fills this gap by postulating a number of axioms for quantifying precision consistently for any log and any model. Further, we show through counter-examples that none of the existing measures consistently quantifies precision.
\end{abstract}


\end{frontmatter}


\section{Introduction}
\label{sec:introduction}
Process mining \cite{Aalst2016} is a fast growing discipline that is focused on the analysis of events logged during the execution of a business process. Events contain information on what was done, by whom, for whom, where, when, etc. Such event data are often readily available from information systems such as ERP, CRM, or BPM systems. Process discovery, which plays a prominent role in process mining, is the task of automatically generating a process model that accurately describes a business process based on such event data. Many process discovery techniques have been developed over the last decade (e.g. \cite{Goedertier2009,Leemans2013,Conforti2016,Zelst2015}), producing process models in various forms, such as Petri nets~\cite{Murata1989}, process trees \cite{Buijs2012}, YAWL models~\cite{Aalst2005}, and BPMN models~\cite{OMG2011}.\looseness=-1

The process model that is pursued by process discovery techniques ideally allows for all the behavior that was observed in the event log (called \emph{fitness}), while at the same time it should not be too general by allowing for much more behavior than what was seen in the event log (called \emph{precision}).

A range of measures have been proposed for quantifying precision \cite{Greco2006,Rozinat2008,Munoz2010,Broucke2014,Leemans2016}. However, to the best of our knowledge, there is currently no work on verifying whether precision measures actually quantify what they are supposed to measure in a consistent manner. Conceptually, the precision of a process model in the context of an event log should be high when the model allows for few traces not seen in the log, and it should be low when it allows for many traces not seen in the log. In this paper we propose a set of axioms that formulate desired properties of precision measures and systematically validate whether these axioms hold for existing precision measures.

In Section \ref{sec:preliminaries} we introduce basic notation and definitions. In Section \ref{sec:axioms} we formulate axioms for precision measures. We then continue with Section \ref{sec:precision_metrics}, where we describe existing precision measures in more detail and validate the axioms for these measures. In Section \ref{sec:undefined_situations} we describe two contexts in which we are not able to define axioms for precision. In Section \ref{sec:conclusion} we conclude this paper and state several directions for future work.
%\noindent\textbf{Soundness} \cite{Greco2006} was the first precision metric and was proposed 2006. Soundness is defined as the percentage of the traces allowed by a process model that are observed in the log. The limitation of soundness is that it is zero for all process models with loops, where the set of traces allowed by the model is infinitely large.

%\noindent\textbf{Behavioral Appropriateness} Rozinat and van der Aalst \cite{Rozinat2008} proposed two precision metrics: Simple Behavioral Appropriateness \cite{Rozinat2008} and Advanced Behavioral Appropriateness. Simple Behavioral Appropriateness measures the amount of possible behavior by determining the mean number of enabled transitions during log replay. Some critique \cite{Weerdt2010} on Simple Behavioral Appropriateness is focused on its property that this metric is dependent on the structural properties of the process model and not solely on the behavior that the model allows. Advanced Behavioral Appropriateness compares the behavior of the model and the event log by analyzing the follows and precedes relations in both. Exhaustive simulation is used to create these relations for the log, which is computationally very demanding, as shown in \cite{Weerdt2010}.

\section{Preliminaries}
\label{sec:preliminaries}
In this section we introduce concepts used in later sections of this paper.

$X=\{a_1,a_2,\dots,a_n\}$ denotes a finite set. $\mathcal{P}(X)$ denotes the power set of $X$, i.e., the set of all possible subsets of $X$. $X^*$ denotes the set of all sequences over a set $X$ and $\sigma=\langle a_1,a_2,\dots,a_n\rangle$ denotes a sequence of length $n$, with $\langle\rangle$ the empty sequence. $X{\setminus}Y$ denotes the set of elements that are in set $X$ but not in set $Y$, e.g., $\{a,b,c\}{\setminus}\{a,c\}{=}\{b\}$. A multiset (or bag) over $X$ is a function $B:X{\rightarrow}\mathbb{N}$ which we write as $[a_1^{w_1},a_2^{w_2},\dots,a_n^{w_n}]$, where for $1{\le} i {\le} n$ we have $a_i{\in} X$ and $w_i{\in}\mathbb{N}^{+}$. The set of all bags over $X$ is denoted $\mathcal{B}(X)$.% and $\sigma_1 \cdot \sigma_2$ is the concatenation of sequences $\sigma_1,\sigma_2$.

In the context of process mining, we assume the set of all \emph{process activities} $\Sigma$ to be given. Event logs consist of sequences of events where each event represents a process activity.
\begin{definition}[Event, Trace, and Event Log]
An \emph{event} $e$ in an event log is the occurrence of an activity $e{\in}\Sigma$. We call a sequence of events $\sigma{\in}\Sigma^*$ a \emph{trace}. An \emph{event log} $L{\in}\mathcal{B}({\Sigma^*})$ is a finite multiset of traces.
\end{definition}
$L{=}[\langle a,b,c\rangle^2,\langle b,a,c\rangle^3]$ is an example event log over process activities $\Sigma{=}\{a,b,c\}$, consisting of 2 occurrences of trace $\langle a,b,c\rangle$ and three occurrences of trace $\langle b,a,c\rangle$.

Most precision measures have been implemented for Petri nets, a process modeling formalism frequently used in the context of process mining. A Petri net is a directed bipartite graph consisting of places (depicted as circles) and transitions (depicted as rectangles), connected by arcs. A transition describes an activity, while places represent the enabling conditions of transitions. Labels of transitions indicate the type of activity that they represent. Unlabeled transitions ($\tau$-transitions) represent invisible transitions (depicted as gray rectangles), which are only used for routing purposes and are not recorded in the event log.
\begin{definition}[Labeled Petri net]
	\label{def:lpn}
	A \emph{labeled Petri net} $N=\langle P,T,F,\ell\rangle$ is a tuple where $P$ is a finite set of places, $T$ is a finite set of transitions such that $P{\cap}T{=}\emptyset$,  $F{\subseteq}(P {\times}T){\cup}(T{\times}P)$ is a set of directed arcs, called the flow relation, and $\ell{:}T{\nrightarrow}\Sigma$ is a partial labeling function that assigns a label to a transition, or leaves it unlabeled (the $\tau$-transitions).\looseness=-1
\end{definition}
We write $\bullet{n}$ and $n\bullet$ for the input and output nodes of $n\in P \cup T$ (according to $F$). A state of a Petri net is defined by its \emph{marking} $m{\in} \mathcal{B}(P)$ being a multiset of places. A marking is graphically denoted by putting $m(p)$ tokens on each place $p{\in}P$. State changes occur through transition firings. A transition $t$ is enabled (can fire) in a given marking $m$ if each input place $p{\in}{\bullet}t$ contains at least one token. Once $t$ fires, one token is removed from each input place $p{\in}{\bullet} t$ and one token is added to each output place $p'{\in}t \bullet$, leading to a new marking $m'{=}m{-}\bullet\!{t}+t\bullet$.

A firing of a transition $t$ leading from marking $m$ to marking $m'$ is denoted as step $m {\step{t}} m'$. Steps are lifted to sequences of firing  enabled transitions, written $m {\step{\gamma}} m'$ and $\gamma {\in}T^*$ is a \emph{firing sequence}..\looseness=-1

A partial function $f{\in} X {\nrightarrow} Y$ with domain $\mathit{dom}(f)$ can be lifted to sequences over $X$ using the following recursive definition: (1) $f(\langle\rangle)=\langle\rangle$;  (2) for any $\sigma{\in} X^*$ and $x\in X$:
\begin{center}
	$f(\sigma \cdot \langle x\rangle) =
	\left\{
	\begin{array}{ll}
	f(\sigma)  & \mbox{if } x{\notin}\mathit{dom}(f), \\
	f(\sigma) \cdot \langle f(x)\rangle & \mbox{if } x{\in}\mathit{dom}(f).
	\end{array}
	\right.$
\end{center}

Defining an \emph{initial} and \emph{final} markings allows to define the \emph{language} accepted by a Petri net as a set of finite sequences of activities.

\begin{definition}[Accepting Petri Net]
	An \emph{accepting Petri net} is a triplet $\APN{=}(N,m_0,\MF)$, where $N$ is a labeled Petri net, $m_0{\in}\mathcal{B}(P)$ is its initial marking, and $\MF{\subseteq}\mathcal{B}(P)$ is its set of possible final markings. A sequence $\sigma{\in}\Sigma^*$ is a \emph{trace} of an accepting Petri net $\APN$ if there exists a firing sequence $m_0{\step{\gamma}}m_f$ such that $m_f{\in}\MF$, $\gamma{\in}T^*$ and $\ell(\gamma){=}\sigma$.
\end{definition}
The \emph{language} $\Lan(\APN)$ is the set of all its traces, i.e., $\Lan(\APN){=}\{l(\gamma)|\gamma{\in}T^*{\land}\exists_{m_f{\in}MF}m_0{\step{\gamma}}m_f\}$, which can be of infinite size when $\APN$ contains loops. Even though we define language for accepting Petri nets, in theory $\Lan(M)$ can be defined for any process model $M$ with formal semantics. We denote the universe of process models as $\mathcal{M}$. For each $M{\in}\mathcal{M}$, $\Lan(M)$ is defined.

For an event log $L$, $\tilde{L}{=}\{\sigma{\in}\Sigma^*|L(\sigma){>}0\}$ is the \emph{trace set} of $L$. For example, for log $L{=}[\langle a,b,c\rangle^2,\langle b,a,c\rangle^3]$, $\tilde{L}{=}\{\langle a,b,c\rangle\langle b,a,c\rangle\}$. For an event log $L$ and a model $M$ we say that $L$ is \emph{fitting} on model $M$ if $\tilde{L}{\subseteq}\Lan(M)$. Precision is related to the behavior that is allowed by a model $M$ that was not observed in the event log $L$, i.e., $\Lan(M){\setminus}\tilde{L}$.%The four models in Fig. \ref{fig:motivating_example} are  accepting Petri nets. Places that belong to the initial marking contain a token and places belonging to a final marking are simply marked as $\begin{tikzpicture}
\section{Axioms for Precision Metrics}
\label{sec:axioms}
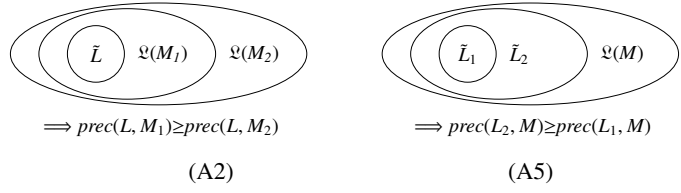
\begin{figure*}
	\renewcommand*\thesubfigure{A\arabic{subfigure}}
	\centering
	\begin{subfigure}{0.33\linewidth}
		\addtocounter{subfigure}{1}
		%\hspace{-0.4cm}
		\scalebox{0.75}{
			\begin{tikzpicture}
			\node[draw,circle,minimum size=1cm,inner sep=0pt,anchor=west] at (1,0) {$\tilde{L}$};
			\node[draw,ellipse,minimum height=1.6cm, minimum width=3.1cm,anchor=west,label={[xshift=-1.5cm]right:$\Lan(\mathit{M_1})$}] at (0.5,0) {};
			\node[draw,ellipse,minimum height=1.8cm, minimum width=5.2cm,anchor=west,label={[xshift=-1.5cm]right:$\Lan(\mathit{M_2})$}] at (0,0) {};
			\node[]at (2.6,-1.3){${\implies}\mathit{prec}(L,M_1){\ge}\mathit{prec}(L,M_2)$};
			\end{tikzpicture}}
		\caption{}
		\label{sfig:monotonicity_axiom}
	\end{subfigure}
	\hspace{-1.4cm}
%	\begin{subfigure}{0.33\linewidth}
%		\centering
%		\scalebox{0.75}{
%			\begin{tikzpicture}
%			\node[draw,circle,minimum size=1cm,inner sep=0pt,anchor=west] at (1,0) {$\tilde{L}$};
%			\node[draw,ellipse,minimum height=1.8cm, minimum width=5.2cm,anchor=west,label={[xshift=-3.2cm]right:$\Lan(\mathit{M_1})=\Lan(\mathit{M_2})$}] at (0,0) {};
%			\node[]at (2.6,-1.3){${\implies}\mathit{prec}(L,M_1){=}\mathit{prec}(L,M_2)$};
%			\end{tikzpicture}}
%		\caption{}
%		\label{sfig:language_equivalence_axiom}
%	\end{subfigure}\\
	\begin{subfigure}{0.33\linewidth}
	\addtocounter{subfigure}{2}
	\hspace{0.6cm}
	\scalebox{0.75}{
		\begin{tikzpicture}
		\node[draw,circle,minimum size=1cm,inner sep=0pt,anchor=west] at (1,0) {$\tilde{L}_1$};
		\node[draw,ellipse,minimum height=1.6cm, minimum width=3.1cm,anchor=west,label={[xshift=-1.5cm]right:$\tilde{L}_2$}] at (0.5,0) {};
		\node[draw,ellipse,minimum height=1.8cm, minimum width=5.2cm,anchor=west,label={[xshift=-1.5cm]right:$\Lan(\mathit{M})$}] at (0,0) {};
		\node[]at (2.6,-1.3){${\implies}\mathit{prec}(L_2,M){\ge}\mathit{prec}(L_1,M)$};
		\end{tikzpicture}}
	\caption{}
	\label{sfig:monotonicity_axiom2}
\end{subfigure}
%	\begin{subfigure}{0.33\linewidth}
	%\hspace{-0.4cm}
%	\scalebox{0.75}{
%		\begin{tikzpicture}
%		\node[draw,ellipse,minimum height=1.6cm, minimum width=3.1cm,anchor=west,label={[xshift=-2.1cm]right:$\tilde{L}_1=\tilde{L}_2$}] at (0.5,0) {};
%		\node[draw,ellipse,minimum height=1.8cm, minimum width=5.2cm,anchor=west,label={[xshift=-1.5cm]right:$\Lan(\mathit{M})$}] at (0,0) {};
%		\node[]at (2.6,-1.3){${\implies}\mathit{prec}(L_1,M){=}\mathit{prec}(L_2,M)$};
%		\end{tikzpicture}}
%	\caption{}
%	\label{sfig:monotonicity_axiom3}
%\end{subfigure}
%\begin{subfigure}{0.33\linewidth}
%	\scalebox{0.75}{
%		\begin{tikzpicture}
%		\node[draw,ellipse,minimum height=1.8cm, minimum width=5.2cm,anchor=west,label={[xshift=-3.2cm]right:$\tilde{L}=\Lan(\mathit{M})$}] at (0,0) {};
%		\node[]at (2.5,-1.3){${\implies}\mathit{prec}(L,M){=}1$};
%		\end{tikzpicture}}
%	\caption{}
%	\label{sfig:unit_precision_axiom}
%\end{subfigure}
	%\begin{subfigure}{0.24\linewidth}
	%	\centering
	%	\scalebox{0.67}{
	%		\begin{tikzpicture}
	%		\node[draw,circle,minimum size=1cm,inner sep=0pt,anchor=west] at (1,0) {$\tilde{L}$};
	%		\node[draw,ellipse,minimum height=1.8cm, minimum width=5.2cm,anchor=west,label={[xshift=-3.2cm]right:$\Lan(\mathit{M})$}] at (0,0) {};
	%		\node[]at (2.6,-1.3){${\implies}\mathit{prec}(L,M){=}\mathit{prec}(L,M)$};
	%		\end{tikzpicture}}
	%	\caption{}
	%	\label{sfig:stability_axiom}
	%\end{subfigure}
	\caption{Two of the five axioms for precision measures visualized with Euler diagrams.}
	\label{fig:precision_axioms}
	\vspace{-0.3cm}
\end{figure*}

The properties that are desired for precision measures are not clearly defined in existing work, although they are often discussed informally. Van der Aalst et al.~\cite{Aalst2011}, describe the precision dimension as ``Precision: measure determining whether the model prohibits behavior very different from the behavior seen in the event log. A model with low precision is “underfitting”.''. Vanden Broucke et al.~\cite{Broucke2014} describe precision as ``precision (or: appropriateness), i.e., the model's ability to disallow unwanted behavior;''. M\~{u}noz-Gama and Carmona~\cite{Munoz2010} describe it as ``Precision: refers to overly general models, preferring models with minimal behavior to represent as closely as possible to the log.''. Buijs et al.~\cite{Buijs2014} describe precision as ``... precision quantifies the fraction of the behavior allowed by the model which is not seen in the event log.''.

We consider precision to be a function $\mathit{prec}(L,M)$ which quantifies which part of the language of model  $M$ is seen in event log $L$. Below we formalize the desired properties of function $\mathit{prec}$ through axioms to \emph{consistently} hold for any kind of model and any kind of log.. Note that in the examples that we will show in this paper all models $M$ will be Petri nets, however the formulated axioms are more general and apply to any process model $M{\in}\mathcal{M}$. Figure \ref{fig:precision_axioms} visualizes two axioms using Euler diagrams.\looseness=-1

%As first axiom we state the precision measure results in the same precision score when precision is calculated repeatedly when calculating the precision for a given log and model.
The first axiom states that precision is deterministic, i.e., given a log and model always the same result is returned.
\begin{axiom}
A precision measure is a \emph{function} $\mathit{prec}:\mathcal{B}(\Sigma^*)\times\mathcal{M}\rightarrow\mathbb{R}$, i.e., it is deterministic.
\end{axiom}

\noindent Existing precision measures normalize $\mathbb{R}$ to a $[0,1]$-interval.

The second axiom formulates the conceptual description of precision more formally: if a process model $M_2$ allows for more behavior not seen in a log $L$ than another model $M_1$ does, then $M_2$ should have a lower precision than $M_1$ regarding $L$.
\begin{axiom}
For models $M_1$ and $M_2$ and a log $L$, $\tilde{L}{\subseteq}\Lan(M_1){\subseteq}\Lan(M_2){\implies}\mathit{prec}(L,M_1){\ge}\mathit{prec}(L,M_2)$
\end{axiom}

Note that \textbf{A2} does allow $\tilde{L}{\subseteq}\Lan(M_1){\subset}\Lan(M_2)$ with $\mathit{prec}(L,M_1){=}\mathit{prec}(L,M_2)$. Ideally, since $\Lan(M_1)$ is smaller than $\Lan(M_2)$ we would like to see a higher precision for $M_1$, but this requirement might be too strict. However, for a process model $M$ with $\tilde{L}{\subseteq}\Lan(M)$, we would like the precision of $M$ on $L$ to be higher than the precision of $M$ on any flower model (i.e., a model that allows for all behavior over its activities) on log $L$.
\begin{axiom}
For models $M_1$ and $M_2$ and a log $L$, $\Lan(M_1){\subset}\mathcal{P}(\Sigma^*){\land}\Lan(M_2){=}\mathcal{P}(\Sigma^*){\implies}\mathit{prec}(L,M_1){>}\mathit{prec}(L,M_2)$
\end{axiom}

The precision of a log on two language equivalent models should be equal, i.e., precision should not depend on the model structure.
\begin{axiom}
For models $M_1$ and $M_2$ and a log $L$, $\Lan(M_1){=}\Lan(M_2){\implies}\mathit{prec}(L,M_1){=}\mathit{prec}(L,M_2)$
\end{axiom}
\textbf{A4} was stated before in an informal manner by Rozinat and van der Aalst \cite{Rozinat2008}, who stated that precision should be independent of structural properties of the model.

Adding fitting traces to a fitting log can only increase the precision of a given model with respect to the log.
\begin{axiom}
For model $M$ and logs $L_1$ and $L_2$, $\tilde{L}_1{\subseteq}\tilde{L}_2{\subseteq}\Lan(M){\implies}\mathit{prec}(L_2,M){\ge}\mathit{prec}(L_1,M)$
\end{axiom}

From \textbf{A5} it follows as a corollary that precision is maximal when the log contains all the traces allowed by the model, and minimal when it contains no traces allowed by the model.

In the coming sections we will validate whether these axioms hold for several precision measures. Some articles that introduce precision measures explicitly mention that the measure is intended to be used only with a certain subclass of Petri nets. An example of such a subclass of Petri nets are bounded Petri nets, which have the restriction that all places most have a finite number of tokens in all reachable markings. When an article that introduces a precision measure states an explicit assumption on the subclass of Petri nets, then we only validate the axioms on this subclass of Petri nets. When no explicit assumption on a subclass of Petri nets is stated, we assume that the precision measure is intended for Petri nets in general.

\section{Precision Metrics}
\label{sec:precision_metrics}
In this section we give an overview of the precision measures that have been developed in the process mining field, and validate the axioms for precision measures introduced in Section \ref{sec:axioms} for each of those measures.

\subsection{Soundness}
Greco et al. \cite{Greco2006} were the first to propose a precision measure, defining it as the number of unique executions of the process that were seen in the event log divided by the number of unique paths through the process model. This measure is not usable in practice, because it is zero when the process model allows for an infinite number of paths through the model. Any process model having a loop has a precision of 0. More recent precision measures are capable of calculating the precision of a model for an event log even when the models allows for infinite behavior.

\subsection{Behavioral Appropriateness}
Rozinat and Van der Aalst \cite{Rozinat2008} proposed the \emph{simple behavioral appropriateness} precision measure, which looks at the average number of enabled transitions during replay. The authors observed themselves that simple behavioral appropriateness is dependent on the structure of the model, and not solely dependent on the behavior that it allows, therefore \textbf{A4} does not hold for this measure. Furthermore, for a process model that contains silent transitions or duplicate labels it is possible that a given trace can be replayed on this model in multiple ways, where the average number of enabled transitions can depend on the chosen replay path through the model. This replay path through the model is chosen arbitrarily from the possible ways in which the trace can be replayed. This shows that \textbf{A1} does not hold for simple behavioral appropriateness, as it is not deterministic.

In the same paper, Rozinat and van der Aalst \cite{Rozinat2008} propose \emph{advanced behavioral appropriateness}, which is independent of the model structure. Advanced behavioral appropriateness calculates the sets $S_{\!F}{\subseteq}\Sigma{\times}\Sigma$ of pairs of activities that sometimes, but not always, follow each other. Likewise set $S_{\!P}{\subseteq}\Sigma{\times}\Sigma$ is calculated as the set of activities that sometimes, but not always, precede each other. $S_{\!F}^{\!L}$ and $S_{\!P}^{\!L}$ denote the sometimes-follows and  sometimes-precedes relations on the log, and $S_{\!F}^{\!M}$ and $S_{\!P}^{\!M}$ denotes the sometimes-follows and sometimes-precedes relations according to the model. However, to calculate $S_{\!F}^{\!M}$ and $S_{\!P}^{\!M}$, exhaustive exploration of the state space of the model is required, prohibiting the application of this measure for large models or highly concurrent models, where the state-space explosion problem arises. Advanced behavioral appropriateness precision is defined as $a'_b{=}(\frac{|S^{\!L}_{\!F}{\cap}S^{\!M}_{\!F}|}{2\cdot|S^{\!M}_{\!F}|}+\frac{|S^{\!L}_{\!P}{\cap}S^{\!M}_{\!P}|}{2\cdot|S^{\!M}_{\!P}|})$. Because $S^{\!M}_{\!F}$ and $S^{\!M}_{\!P}$ are obtained through exhaustive exploration of the state space of the model, it is easy to see that they depend only on the behavior of the model and not on its structure, therefore \textbf{A4} holds. A problem with advanced behavioral appropriateness occurs for deterministic models, where $|S^{\!M}_{\!P}|{=}|S^{\!M}_{\!F}|{=}0$, leading to undefined precision. This shows that advanced behavioral appropriateness is a partial function, which is in conflict with \textbf{A1}.\looseness=-1

\begin{figure}[t]
	\centering
	\begin{tikzpicture}
	[node distance=0.7cm,
	on grid,>=stealth',
	bend angle=20,
	auto,
	every place/.style= {minimum size=4mm},
	every transition/.style = {minimum size = 3mm}
	]
	\node [place, tokens = 1] at (-1.4,0) (p0){};
	\node [transition] (ts4) [minimum width=3mm,fill=lightgray] at (-0.7,0) {}
	edge[pre] node[auto] {} (p0);
	\node [place] at (0,0) (p1){}
	edge[pre] node[auto] {} (ts4);
	\node [place] at (1.4,0) (p2){};
	\node [place] at (2.8,0) (p3) {};
	\node [transition] (t1) [] at (0.7,0.35){a} 
	edge[pre] node[auto] {} (p1)
	edge[post] node[auto] {} (p2);
	\node [transition] (t4) [] at (0.7,-0.35){b} 
	edge[pre] node[auto] {} (p1)
	edge[post] node[auto] {} (p2);
	
	\node [transition] (t2) [align=center] at (2.1,0.35) {c} 
	edge [pre] node[auto] {} (p2)
	edge[post] node[auto] {} (p3);
	\node [transition] (t3) [] at (2.1,-0.35){d} 
	edge[pre] node[auto] {} (p2)
	edge[post] node[auto] {} (p3);
	
	\node [transition] (ts3) [minimum width=3mm,fill=lightgray] at (2.8,0.7) {}
	edge[pre] node[auto] {} (p3);
	\draw [post] (ts3) to [out=180,in=0] ($(t1)+(0,0.4)$) to [out=180,in=90 ] (p1);
	\node [transition] (ts5) [minimum width=3mm,fill=lightgray] at (3.5,0.0) {}
	edge[pre] node[auto] {} (p3);
	\node [place,pattern=custom north west lines,hatchspread=1.5pt,hatchthickness=0.25pt,hatchcolor=gray] at (4.2,0) (p4) {}
	edge[pre] node[auto] {} (ts5);
	\end{tikzpicture}
	\caption{Model $M$.}
	\label{fig:aba_a3_counter_example}
\end{figure}

Rozinat and van der Aalst \cite{Rozinat2008} state that simple behavioral appropriateness and advanced behavioral appropriateness assume the Petri net to be in the class of \emph{sound workflow (WF) nets} \cite{Aalst1997}. A WF-net requires the Petri net to have (i) a single \emph{Start} place, (ii) a single \emph{End} place, and (iii) every node must be on some path from \emph{Start} to \emph{End}. The \emph{soundness} property additionally require that each transition can be potentially executed, and that the process can always terminate properly, i.e., finish with only one token in the \emph{End} place.

Consider model $M$ of Figure \ref{fig:aba_a3_counter_example}, which belongs to the class of sound WF-nets, and any log $L$ such that $\tilde{L}{\subseteq}\Lan(M)$. The loop in model $M$ causes $S^{\!M}_{\!F}$ and $S^{\!M}_{\!P}$ to contain all pairs of activities of $\Sigma$. Therefore, $|S^{\!M}_{\!F}|$ and $|S^{\!M}_{\!P}|$ are identical to the sometimes relations $|S^{\!M'}_{\!F}|$ and $|S^{\!M'}_{\!P}|$ of any model $M'$ with $\Lan(M'){=}\mathcal{P}(\Sigma^*)$, leading to $\mathit{prec}(L,M){=}\mathit{prec}(L,M')$. As $\Lan(M){\subset}\mathcal{P}(\Sigma^*)$, this is in conflict with \textbf{A3}.

\subsection{Escaping Edges Precision}
\begin{figure}
	\begin{subfigure}{0.45\linewidth}
		\centering
		\scalebox{0.9}{
			\begin{tikzpicture}
			\centering
			[node distance=0.7cm,
			on grid,
			bend angle=20,
			every place/.style= {minimum size=4mm},
			every transition/.style = {minimum size = 3mm}
			]
			\node [place,tokens=1](p1) at (0,0) [] {}; %: $p_{1}$ 
			\node [place,pattern=custom north west lines,hatchspread=1.5pt,hatchthickness=0.25pt,hatchcolor=gray](p2) at (2.8,0) [] {}; %: $p_{2}$ 
			\node [place](p3) at (1.4,0) [] {}; %: $p_{3}$ 
			\node [transition](15) at (0.7,0.5) {a};
			\node [transition](2) at (0.7,-0.5) {b};
			\node [transition](3) at (2.1,0.5) {c};
			\node [transition](4) at (2.1,-0.5) {d};
			\draw 
			(2) edge[->,>=stealth] (p1)
			(p1) edge[->,>=stealth] (15)
			(15) edge[->,>=stealth] (p3)
			(p3) edge[->,>=stealth] (2)
			(p3) edge[->,>=stealth] (3)
			(p3) edge[->,>=stealth] (4)
			(3) edge[->,>=stealth] (p2)
			(4) edge[->,>=stealth] (p2);
			\end{tikzpicture}}
		\caption{}
		\label{sfig:loop-choice_model}
	\end{subfigure}
	\begin{subfigure}{0.5\linewidth}
		\centering
		\scalebox{0.9}{
			\begin{tikzpicture}
			\tikzstyle{vertex}=[circle,fill=black!25,minimum size=10pt,inner sep=0pt]
			\node[vertex] (G1) at (0,0) {2};
			\node[vertex] (G2) at (0.7,0) {2};
			\node[vertex] (G4) at (1.4,0.4) {1};
			\node[vertex] (G5) at (1.4,-0.4) {1};
			
			\coordinate (C1) at (1.4,0);
			
			\draw[->]  (G1) to node [auto,above] {a} (G2);
			\draw[->]  (G2) to node [auto,above] {d} (G4);
			\draw[->]  (G2) to node [auto,below] {c} (G5);
			\draw[->,red]  (G2) to node [auto,right,pos=1,red] {b} (C1);
			\end{tikzpicture}}
		\caption{}
		\label{sfig:prefix_automaton_l1}
	\end{subfigure}
	\begin{subfigure}{\linewidth}
		\scalebox{0.9}{
			\begin{tikzpicture}
			\tikzstyle{vertex}=[circle,fill=black!25,minimum size=10pt,inner sep=0pt]
			\node[vertex] (G1) at (0,0) {3};
			\node[vertex] (G2) at (0.7,0) {3};
			\node[vertex] (G3) at (1.4,0) {1};
			\node[vertex] (G4) at (1.4,0.8) {1};
			\node[vertex] (G5) at (1.4,-0.8) {1};
			
			\node[vertex] (G6) at (2.1,0) {1};
			\node[vertex] (G7) at (2.8,0) {1};
			\node[vertex] (G8) at (3.5,0) {1};
			\node[vertex] (G9) at (4.2,0) {1};
			\node[vertex] (G10) at (4.9,0) {1};
			\node[vertex] (G11) at (5.6,0) {1};
			\node[vertex] (G12) at (6.3,0) {1};
			\node[vertex] (G13) at (7.0,0) {1};
			\node[vertex] (G14) at (7.7,0) {1};
			\node[vertex] (G15) at (8.4,0) {1};
			
			\coordinate (C1) at (2.5,0.7);
			\coordinate (C2) at (2.5,-0.7);
			
			\coordinate (C3) at (3.9,0.7);
			\coordinate (C4) at (3.9,-0.7);
			
			\coordinate (C5) at (5.3,0.7);
			\coordinate (C6) at (5.3,-0.7);
			
			\coordinate (C7) at (6.7,0.7);
			\coordinate (C8) at (6.7,-0.7);
			
			\coordinate (C9) at (8.1,0.7);
			\coordinate (C10) at (8.1,-0.7);
			
			\draw[->]  (G1) to node [auto,above] {a} (G2);
			\draw[->]  (G2) to node [auto,above,pos=0.8] {b} (G3);
			\draw[->]  (G2) to node [auto,above] {d} (G4);
			\draw[->]  (G2) to node [auto,below] {c} (G5);
			
			\draw[->]  (G3) to node [auto,above] {a} (G6);
			\draw[->,red]  (G6) to node [auto,above,pos=0.4,red] {d} (C1);
			\draw[->,red]  (G6) to node [auto,above,pos=0.8,red] {a} (C2);
			
			\draw[->]  (G6) to node [auto,above] {b} (G7);
			\draw[->]  (G7) to node [auto,above] {a} (G8);
			\draw[->,red]  (G8) to node [auto,above,pos=0.4,red] {d} (C3);
			\draw[->,red]  (G8) to node [auto,above,pos=0.8,red] {c} (C4);
			
			\draw[->]  (G8) to node [auto,above] {b} (G9);
			\draw[->]  (G9) to node [auto,above] {a} (G10);
			\draw[->,red]  (G10) to node [auto,above,pos=0.4,red] {d} (C5);
			\draw[->,red]  (G10) to node [auto,above,pos=0.8,red] {c} (C6);
			
			\draw[->]  (G10) to node [auto,above] {b} (G11);
			\draw[->]  (G11) to node [auto,above] {a} (G12);
			\draw[->,red]  (G12) to node [auto,above,pos=0.4,red] {d} (C7);
			\draw[->,red]  (G12) to node [auto,above,pos=0.8,red] {c} (C8);
			
			\draw[->]  (G12) to node [auto,above] {b} (G13);
			\draw[->]  (G13) to node [auto,above] {a} (G14);
			\draw[->,red]  (G14) to node [auto,above,pos=0.4,red] {d} (C9);
			\draw[->,red]  (G14) to node [auto,above,pos=0.8,red] {b} (C10);
			\draw[->]  (G14) to node [auto,above] {c} (G15);
			\end{tikzpicture}}
		\caption{}
		\label{sfig:prefix_automaton_l2}
	\end{subfigure}
	
	\caption{\emph{(a)} Model $M$, and the alignment automata on Model $M$ for \emph{(b)} log $L_1{=}[\langle a,c\rangle,\langle a,d \rangle]$, and for \emph{(c)} log $L_2{=}[\langle a,c\rangle,\langle a,d\rangle,\langle a,b,a,b,a,b,a,b,a,b,a,c\rangle]$. Red arcs correspond to escaping edges.}
	\label{fig:prefix_automaton}
\end{figure}
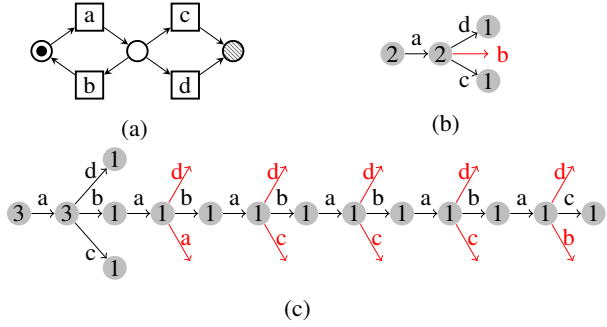

Escaping Edges Precision (ETC) \cite{Munoz2010} calculates precision by constructing a \emph{prefix automaton}, which consists of one state per unique prefix of the event log. Figure \ref{sfig:prefix_automaton_l1} shows an example prefix automaton for an event log $L=[\langle a,c\rangle,\langle a,d\rangle]$. For each state in the prefix automaton it is then determined which activities are allowed as next activities by the process model. Activities that are allowed as next activities for some prefix but that are never observed in the event log after this prefix are referred to as \emph{escaping edges}.

In later work \cite{Adriansyah2012,Aalst2012}, alignments \cite{Adriansyah2011} are used to calculate the prefix automaton on the aligned event log instead of the original event log, making the precision measure robust to non-fitting traces, i.e., traces that are not in the language of the model. For a trace $\sigma$ from a log $L$ that is fitting on an accepting Petri net $\mathit{APN}$, alignments \cite{Adriansyah2011} give a sequence of transition firings $\gamma{\in}T^*$ such that $m_0{\step{\gamma}}m_f$ with $m_0$ the initial marking and $m_f$ a final marking of $\mathit{APN}$ and $\ell(\gamma){=}\sigma$. Note that for a given trace $\sigma$ and model, multiple possible alignments can exist. For non-fitting traces, alignments search for a firing sequence $\gamma{\in}T^*$ such that $\ell(\gamma)$ is as close as possible to $\sigma$. Adriansyah et al. \cite{Adriansyah2012} describe two versions of the alignment-based escaping edges precision: \emph{one-align ETC}, which calculates the precision based on one optimal alignment of log and model, and \emph{all-align ETC}, which calculates the precision based on all optimal alignments between log and model. In practice, it is often computationally infeasible to calculate all optimal alignments. A later precision measure, \emph{representative-align ETC} \cite{Adriansyah2015}, calculates the escaping edges based on a sample of optimal alignments, and can therefore be seen as a trade-off between the computational efficiency of one-align ETC and the reliability of all-align ETC. The papers on ETC precision and its variants do not state an assumption on a subclass of Petri nets. ETC, one-align ETC, all-align, and representative-align ETC precision are all implemented in the package ETConformance\footnote{https://svn.win.tue.nl/trac/prom/browser/Packages/ETConformance} as part of the process mining framework ProM~\cite{Dongen2005}.\looseness=-1

The one optimal alignment that is used by one-align ETC is chosen arbitrarily from the set of optimal alignments of a log on a model. However, different optimal alignments result in different prefix automata, which can potentially lead to different precision values. This shows that \textbf{A1} does not hold for one-align ETC.\looseness=-1

Consider log $L_1{=}[\langle a,c\rangle,\langle a,d\rangle]$, log $L_2{=}[\langle a,c\rangle,\langle a,d\rangle,\allowbreak\langle a,c\rangle,\langle a,b,a,b,a,b,a,b,a,b,a,c\rangle]$ and model $M$ be the Petri net of Figure \ref{sfig:loop-choice_model}. Note that $\tilde{L}_1{\subset}\tilde{L}_2$. The alignment automata generated for the calculation of $\mathit{prec}(L_1,M)$ and the calculation of $\mathit{prec}(L_2,M)$ are shown in Figure \ref{sfig:prefix_automaton_l1} and Figure \ref{sfig:prefix_automaton_l2}. The circles represent the states of the automaton, and the arrows the transitions. The numbers in the states represent the weights of the states for the precision calculation, i.e., the number of times that states are visited in the alignment of log $L$ on model $M$ \cite{Adriansyah2012}. In an alternative definition of one-align ETC \cite{Aalst2012} the states are weighted by the number of times that events occurred while being in this state according to the alignment of $L$ on $M$, instead of the number of times that this state was reached according to this alignment. Figure \ref{sfig:prefix_automaton_l1} shows that the initial state was visited twice, activity $a$ occurred twice at the start in log $L_1$, resulting in a state from which activities $b$, $d$, and $c$ were enabled. From this state, activities $c$ and $d$ were seen once, and activity $b$ was never seen, thus it is an escaping edge. Escaping edges precision is then the (weighted) average ratio of non-escaping edges from all outgoing edges, where states are weighted by the number of times that they are visited. Counting the weighted number of non-escaping edges in the numerator and the weighted total number of edges in the denominator in our example, we find $\mathit{prec}(L_1,M){=}\frac{2{\times}1+2{\times}2+1{\times}0+1{\times}0}{2{\times}1+2{\times}3+1{\times}0+1{\times}0}{=}\frac{6}{8}{=}0.75$. One-align ETC results in the following precision values for $M$ on $L_1$ and $L_2$: $\mathit{prec}(L_1,M){=}0.75$ and $\mathit{prec}(L_2,M){=}0.7143$. This shows that \textbf{A5} does not hold for one-align ETC. By comparing the automata of Figures \ref{sfig:prefix_automaton_l1} and \ref{sfig:prefix_automaton_l2} it becomes clear that the single trace that is in $L_2$ but not in $L_1$ brings the model to many states with three escaping edges, reducing precision. The prefix automata and the precision calculations for $M$ on logs $L_1$ and $L_2$ were performed manually following the procedure from the paper and validation using the ETConformance plugin in ProM.

\begin{figure}[t]
	\begin{subfigure}{0.37\linewidth}
		\centering
		\begin{tikzpicture}
		[node distance=0.7cm,
		on grid,>=stealth',
		bend angle=20,
		auto,
		every place/.style= {minimum size=4mm},
		every transition/.style = {minimum size = 3mm}
		]
		\node [place, tokens = 1] at (0,0) (p1){};
		\node [place,pattern=custom north west lines,hatchspread=1.5pt,hatchthickness=0.25pt,hatchcolor=gray] (p2) [] at (1.4,0) {};
		\node [transition] (t1) [align=center] at (0.7,1.05) {a} 
		edge [pre] node[auto] {} (p1)
		edge[post] node[auto] {} (p2);
		\node [transition] (t2) [] at (0.7,0.35){b} 
		edge[pre] node[auto] {} (p1)
		edge[post] node[auto] {} (p2);
		\node [transition] (t3) [] at (0.7,-0.35) {c} 
		edge[pre] node[auto] {} (p1)
		edge[post] node[auto] {} (p2);
		\node [transition] (ts1) [minimum width=3mm,fill=lightgray] at (0.7,-1.05){}
		edge[pre] node[auto] {} (p1)
		edge[post] node[auto] {} (p2);
		\node [transition] (ts2) [minimum width=3mm,fill=lightgray] at (2.1,0) {}
		edge[pre] node[auto] {} (p2);
		\draw [post] (ts2) to [out=90,in=0] ($(t1)+(0,0.5)$) to [out=180,in=90 ] (p1);
		\end{tikzpicture}
		\caption{}
		\label{sfig:traditional_flower}
	\end{subfigure}
	\begin{subfigure}{.32\linewidth}
		\centering
		\begin{tikzpicture}
		[node distance=0.7cm,
		on grid,>=stealth',
		bend angle=20,
		auto,
		every place/.style= {minimum size=4mm},
		every transition/.style = {minimum size = 3mm}
		]
		\node [place, tokens = 1] at (0,0) (p1){};
		\node [place] at (1.2,0) (p2){};
		\node [place] at (2.4,0) (p3){};
		\node [place,pattern=custom north west lines,hatchspread=1.5pt,hatchthickness=0.25pt,hatchcolor=gray] at (3.6,0) (p4) {};
		\node [transition] (t1) [] at (0.6,0.3){c} 
		edge[pre] node[auto] {} (p1)
		edge[post] node[auto] {} (p2);
		\node [transition] (ts1) [minimum width=3mm,fill=lightgray] at (0.6,-0.3){}
		edge[pre] node[auto] {} (p1)
		edge[post] node[auto] {} (p2);
		
		\node [transition] (t2) [align=center] at (1.8,0.3) {b} 
		edge [pre] node[auto] {} (p2)
		edge[post] node[auto] {} (p3);
		\node [transition] (ts2) [minimum width=3mm,fill=lightgray] at (1.8,-0.3){}
		edge[pre] node[auto] {} (p2)
		edge[post] node[auto] {} (p3);
		
		\node [transition] (t3) [align=center] at (3.0,0.3) {a} 
		edge [pre] node[auto] {} (p3)
		edge[post] node[auto] {} (p4);
		\node [transition] (ts2) [minimum width=3mm,fill=lightgray] at (3.0,-0.3){}
		edge[pre] node[auto] {} (p3)
		edge[post] node[auto] {} (p4);
		
		\node [transition] (ts4) [minimum width=3mm,fill=lightgray] at (4.2,0) {}
		edge[pre] node[auto] {} (p4);
		\draw [post] (ts4) to [out=90,in=0] ($(t2)+(0,0.5)$) to [out=180,in=90 ] (p1);
		\end{tikzpicture}
		\caption{}
		\label{sfig:skip_flower}
	\end{subfigure}\\
	\begin{subfigure}{\linewidth}
		\centering
		\begin{tikzpicture}
		[node distance=0.7cm,
		on grid,>=stealth',
		bend angle=20,
		auto,
		every place/.style= {minimum size=4mm},
		every transition/.style = {minimum size = 3mm}
		]
		\node [place, tokens = 1] at (0,0) (p1){};
		\node [place] at (1.4,0) (p2){};
		\node [place,pattern=custom north west lines,hatchspread=1.5pt,hatchthickness=0.25pt,hatchcolor=gray] at (2.8,0) (p3) {};
		\node [transition] (t1) [] at (0.7,0.0){a} 
		edge[pre] node[auto] {} (p1)
		edge[post] node[auto] {} (p2);
		
		\node [transition] (t2) [align=center] at (2.1,1.05) {b} 
		edge [pre] node[auto] {} (p2)
		edge[post] node[auto] {} (p3);
		\node [transition] (t3) [] at (2.1,0.35){c} 
		edge[pre] node[auto] {} (p2)
		edge[post] node[auto] {} (p3);
		\node [transition] (ts1) [minimum width=3mm,fill=lightgray] at (2.1,-0.35){}
		edge[pre] node[auto] {} (p2)
		edge[post] node[auto] {} (p3);
		
		\node [transition] (ts3) [minimum width=3mm,fill=lightgray] at (3.5,0) {}
		edge[pre] node[auto] {} (p3);
		\draw [post] (ts3) to [out=90,in=0] ($(t2)+(0,0.4)$) to [out=180,in=90 ] (p2);
		\end{tikzpicture}
		\caption{}
		\label{sfig:non-flower}
	\end{subfigure}
	\caption{\emph{(a)} The flower-model, which allows for all behavior over its set of activities, \emph{(b)} an alternative representation of the flower-model, and \emph{(c)} a constrained model which always starts with activity $a$.}
	\vspace{-0.25cm}
\end{figure}
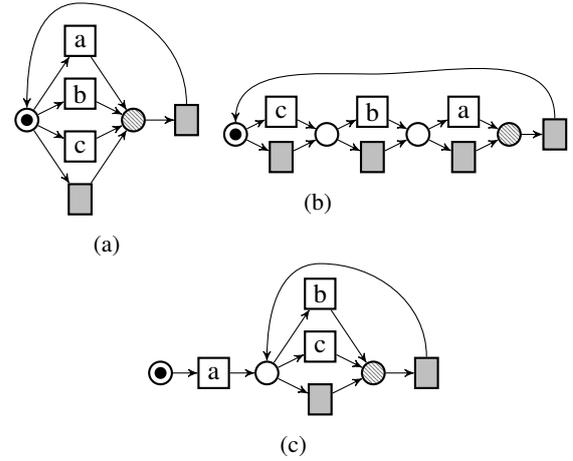

Now consider log $L=[\langle a,b,c\rangle]$, and the three Petri nets $M_1$, $M_2$, $M_3$ in Figures \ref{sfig:traditional_flower}, \ref{sfig:skip_flower}, and \ref{sfig:non-flower} respectively. Note that $M_1$ and $M_2$ are language equivalent, as $\Lan(M_1){=}\Lan(M_2){=}\{a,b,c\}^*$. $M_3$ is more behaviorally constrained than $M_1$ and $M_2$, since all its traces start with activity $a$. The one-align precision of $M_1$, $M_2$, $M_3$ on $L$ are: $\mathit{prec}(L,M_1){=}0.3333$, $\mathit{prec}(L,M_2){=}0.5238$, and $\mathit{prec}(L,M_3){=}0.4444$. $\Lan(M_3){\subset}\Lan(M_1)$, but $\mathit{prec}(L,M_3){>}\mathit{prec}(L,M_1)$, implying that \textbf{A2} does not hold for one-align ETC. Furthermore, $\Lan(M_1){=}\Lan(M_2)$, but $\mathit{prec}(L,M_1){\ne}\mathit{prec}(L,M_2)$, implying that \textbf{A4} does not hold for one-align ETC.

Analyzing the ETConformance plugin in ProM we found that the prefix automaton generated for one-align precision for calculation of $\mathit{prec}(L,M_1)$ results in 6 states, belonging to 3 firings of observable transitions and $2$ firings of $\tau$-transitions. In 3 of the 6 states, which correspond to $M_1$ being in the initial marking, there are 4 possible next activities according to the model, of which only one is observed for that prefix. Furthermore, it shows that the alignment automaton generated for $L$ and $M_1$ consists of 6 states, the automaton for $L$ and $M_2$ consists of 12 states, and the automaton for $L$ and $M_3$ consists of 5 states. This shows that the silent ($\tau$) transitions in $M_2$ generate additional states in the alignment automaton, leading to a higher precision value.

Computing the precision of $M_1$ or of $M_2$ on $L$ did not finish with all-align ETC and representative-align ETC after 8 hours of computation time. The long computation time of all-align ETC and representative-align ETC on models where many optimal alignments exist is a known issue which hinders the application of those measures in practice.

\subsection{Negative Event Precision}

Goedertier et al. \cite{Goedertier2009} proposed a method to induce \emph{negative events}, i.e., sets of events that were prevented from taking place. Negative events are induced for each position in the event log, i.e., for each event $e$ in each trace of the log a set of events is induced that could not have taken place instead of event $e$. De Weerdt et al. \cite{Weerdt2011} proposed a precision measure based on negative events, \emph{behavioral precision} ($p_B$), which is closely linked to how precision is defined in the area of data mining. Negative event precision regards a process model as a binary classifier that determines whether a certain event can take place given a certain prefix, and then evaluates the precision of this classifier in data mining terms taking the induced negative events as ground truth. For a given trace prefix, \emph{true positive} (TP) events are defined as events that are possible according to both the process model (i.e., a transition labeled with this event is enabled) and log (i.e., this event is not a negative event). \emph{False positive} events (FP) are negative events induced for a given prefix that were possible according to the model. Behavioral precision is defined as $p_B=\frac{TP}{TP+FP}$, which is in accordance to the definition of precision in the data mining field. In later work \cite{Broucke2012} induction of artificial negative events has been refined based on frequent temporal patterns which are mined from the event log. Finally, weighted artificial events, where negative events are weighted according to their confidence, are proposed in \cite{Broucke2014}. 

\emph{Weighted behavioral precision} induces negative events for an event $e$ in the log by taking a window of events $w$ that directly precede $e$, then calculating all subsequences of events in the log that exactly match $w$, and finally negative events are identified by calculating which events have never occurred in the log directly after any subsequence matching $w$. This procedure is repeated for different windows sizes, and the resulting negative event are weighted by window size.

To induce the events that could not have happened after e.g. trace prefix $\sigma'=\langle a,c,c,d,e,c,d,e,e\rangle,$ the method to induce weighted negative events described in \cite{Broucke2014} searches for subsequences of events in the log that are identical to the latest $k$ events of $\sigma'$ in the event log. All the activities that have never succeeded such subsequences are considered to be negative events, furthermore, the confidence of these negative events is based on the length $k$ of those matching subsequences.

Negative event based precision measures, with the different methods for negative event induction, are implemented in the ProM package NEConformance\footnote{http://processmining.be/neconformance/}. In this paper we evaluate the precision measure that uses weighted negative events \cite{Broucke2014}, which is the most recent approach to induce negative events and the recommended approach for measuring precision \cite{Broucke2014}. No assumption on a subclass of Petri nets is stated in the papers on negative event precision.

\begin{figure}
	\centering
	%\begin{subfigure}{.59\linewidth}
	%\centering
	\scalebox{0.8}{
		\begin{tikzpicture}
		[node distance=0.7cm,
		on grid,>=stealth',
		bend angle=20,
		auto,
		every place/.style= {minimum size=4mm},
		every transition/.style = {minimum size = 3mm}
		]
		\node [place, tokens = 1] at (0,0) (p1){};
		\node [place] at (1.4,0) (p2){};
		\node [place] at (2.8,0) (p3){};
		
		\node [place] at (4.2,0) (p4) {};
		\node [place] at (2.8,1.1) (p5){};
		\node [place] at (2.8,-1.1) (p6){};
		\node [place] at (4.2,1.1) (p7) {};
		\node [place] at (4.2,-1.1) (p8) {};
		\node [place] at (5.6,0) (p9) {};
		\node [transition] (t1) [] at (0.7,0.3){a} 
		edge[pre] node[auto] {} (p1)
		edge[post] node[auto] {} (p2);
		\node [transition] (t4) [minimum width=3mm] at (0.7,-0.3){b}
		edge[pre] node[auto] {} (p1)
		edge[post] node[auto] {} (p2);
		
		\node [transition] (ts2) [minimum width=3mm,fill=lightgray] at (2.1,0){}
		edge[pre] node[auto] {} (p2)
		edge[post] node[auto] {} (p3)
		edge[post] node[auto] {} (p5)
		edge[post] node[auto] {} (p6);
		
		\node [transition] (t3) [align=center] at (3.5,1.4) {c} 
		edge [pre] node[auto] {} (p5)
		edge[post] node[auto] {} (p7);
		\node [transition] (ts3) [minimum width=3mm,fill=lightgray] at (3.5,0.8){}
		edge[pre] node[auto] {} (p7)
		edge[post] node[auto] {} (p5);
		
		\node [transition] (t5) [align=center] at (3.5,0.3) {d} 
		edge [pre] node[auto] {} (p3)
		edge[post] node[auto] {} (p4);
		\node [transition] (ts4) [minimum width=3mm,fill=lightgray] at (3.5,-0.3){}
		edge[pre] node[auto] {} (p4)
		edge[post] node[auto] {} (p3);
		
		\node [transition] (t6) [align=center] at (3.5,-0.8) {e} 
		edge [pre] node[auto] {} (p6)
		edge[post] node[auto] {} (p8);
		\node [transition] (ts5) [minimum width=3mm,fill=lightgray] at (3.5,-1.4){}
		edge[pre] node[auto] {} (p8)
		edge[post] node[auto] {} (p6);
		
		\node [transition] (ts6) [minimum width=3mm,fill=lightgray] at (4.9,0) {}
		edge[pre] node[auto] {} (p4)
		edge[pre] node[auto] {} (p7)
		edge[pre] node[auto] {} (p8)
		edge[post] node[auto] {} (p9);
		
		\node [transition] (t7) [align=center] at (6.3,0.35) {f} 
		edge [pre] node[auto] {} (p9);
		\node [transition] (t8) [align=center] at (6.3,-0.35) {g} 
		edge [pre] node[auto] {} (p9);
		
		\node [place,pattern=custom north west lines,hatchspread=1.5pt,hatchthickness=0.25pt,hatchcolor=gray] at (7.0,0) (p10) {}
		edge [pre] node[auto] {} (t7)
		edge [pre] node[auto] {} (t8);
		
		\node [place,style=dotted] at (3.5,2) (p11){}
		edge [pre, style=dotted, bend right] node[auto] {} (t1)
		edge [post, style=dotted, bend left] node[auto] {} (t7);
		
		\node [place,style=dotted] at (3.5,-2) (p12){}
		edge [pre, style=dotted, bend left] node[auto] {} (t4)
		edge [post, style=dotted, bend right] node[auto] {} (t8);
		\end{tikzpicture}
	}
	\caption{A process model $M_1$ and its behaviorally restricted variant $M_2$ which includes the dotted places and arcs.}
	\label{fig:negative_event_example}
	\vspace{-0.3cm}
\end{figure}
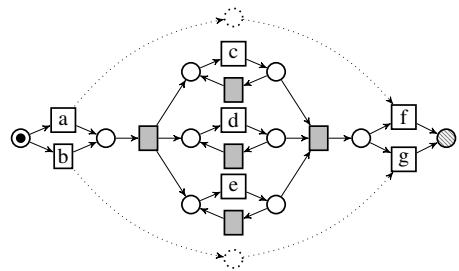

Consider models $M_1$ and $M_2$ of Figure \ref{fig:negative_event_example} respectively excluding and including the arcs and places indicated in dotted lines. $\Lan(M_2){\subset}\Lan(M_1)$, since $M_2$ contains a long term dependency between activities $a$ and $f$ and between activities $b$ and $g$, which $M_1$ does not have. Consider an event log $L$ which consists of $10$ traces from $M_2$, leading to $L$ being fitting on both $M_1$ and $M_2$. We found the negative event precision of $M_1$ and $M_2$ on the same $L$ to be non-deterministic, resulting in slightly different values every time that it is calculated. This shows that \textbf{A1} does not hold for negative event precision. 

Because negative event precision is non-deterministic, we calculated the precision of $M_1$ and $M_2$ on $L$ both $20$ times. The highest precision found in $20$ repetitions for $M_1$ is $0.4876$, while the lowest precision found for $M_2$ is $0.4545$, showing that the non-determinism has the effect that \textbf{A2} does not hold for negative event precision. We found an average value of $0.4744$ with a standard deviation of $0.0090$ for the precision of $M_1$ on $L$ and an average value of $0.4640$ with a standard deviation of $0.0072$ for the precision of $M_2$ on $L$. This shows that also in terms of average precision value \textbf{A2} does not hold. 

To test whether the difference in mean precision between $M_1$ and $M_2$ is due to chance alone we formulate a null hypothesis:\\
$H_0:$ \emph{The average negative event precision of $M_2$ on $L$ is higher than or equal to the average negative event precision of $M_1$ on $L$.}

Testing this null hypothesis with a one-tailed Welch t-test \cite{Welch1947} we found a \emph{p-value} of $0.0001801$, indicating that we can reject the null hypothesis with significance level $0.01$. This shows that, with statistical significance, the precision of $M_1$ on $L$ is higher than the precision of $M_2$ on $L$, which is in disagreement with \textbf{A2}.

To see why \textbf{A2} does not hold for negative event precision, consider the negative event inducing procure being applied to trace prefix $\sigma'=\langle a,c,c,d,e,c,d,e,e\rangle$ from log $L$. Petri net $M_2$ generates many different traces because of the parallel length-one-loops on activities $c$, $d$ and $e$, which allows for any sequence of any length over these activities. Therefore, the matching subsequences of $\sigma'$ in the log generated from $M_2$ are the subsequences that by chance ended in the same behavior over $c$, $d$, and $e$. Because the sequences of $c$, $d$ and $e$ events can be long and diverse, activity $a$ and $b$ are unlikely to be present in the matching subsequences, which makes it unlikely that the procedure can induce the negative event $g$ for $\sigma'$. Because the negative events that reflect the constraint that $M_2$ introduces compared to $M_1$ cannot be induced from the log, negative event precision is not able to recognize that $M_2$ is more precise than $M_1$.

\subsection{Projected Conformance Checking}
\begin{figure}
	\begin{subfigure}{0.3\linewidth}
		\centering
		\begin{tikzpicture}
		[node distance=0.7cm,
		on grid,>=stealth',
		bend angle=20,
		auto,
		every place/.style= {minimum size=4mm},
		every transition/.style = {minimum size = 3mm}
		]
		\node [place, tokens = 1] at (0,0) (p1){};
		\node [place,pattern=custom north west lines,hatchspread=1.5pt,hatchthickness=0.25pt,hatchcolor=gray] (p2) [] at (1.4,0) {};
		\node [transition] (t1) [align=center] at (0.0,1.05) {a} 
		edge [pre] node[auto] {} (p1)
		edge[post] node[auto] {} (p1);
		\node [transition] (t1) [align=center] at (0.7,0) {b} 
		edge [pre] node[auto] {} (p1)
		edge[post] node[auto] {} (p2);
		\end{tikzpicture}
		\caption{}
		\label{sfig:l1l}
	\end{subfigure}
	\quad
	\begin{subfigure}{0.4\textwidth}
		\centering
		\vspace{0.53cm}
		\begin{tikzpicture}
		[node distance=0.7cm,
		on grid,>=stealth',
		bend angle=20,
		auto,
		every place/.style= {minimum size=4mm},
		every transition/.style = {minimum size = 3mm}
		]
		\node [place, tokens = 1] at (0,0) (p1){};
		\node [place] at (1.4,0) (p2){};
		\node [place] at (2.8,0) (p3){};
		\node [place,pattern=custom north west lines,hatchspread=1.5pt,hatchthickness=0.25pt,hatchcolor=gray] at (1.4,1.4) (p4) {};
		\node [transition] (t1) [align=center] at (0.7,0.0) {a} 
		edge [pre] node[auto] {} (p1)
		edge[post] node[auto] {} (p2);
		\node [transition] (t2) [align=center] at (2.1,0.0) {a} 
		edge [pre] node[auto] {} (p2)
		edge[post] node[auto] {} (p3);
		\node [transition] (t3) [align=center] at (0,0.7) {b} 
		edge [pre] node[auto] {} (p1)
		edge[post] node[auto] {} (p4);
		\node [transition] (t4) [align=center] at (1.4,0.7) {b} 
		edge [pre] node[auto] {} (p2)
		edge[post] node[auto] {} (p4);
		\node [transition] (t5) [align=center] at (2.8,0.7) {b} 
		edge [pre] node[auto] {} (p3)
		edge[post] node[auto] {} (p4);
		\end{tikzpicture}
		\caption{}
		\label{sfig:l1l_unrolled}
	\end{subfigure}
	\caption{\emph{(a)} A model with a length-one-loop, \emph{(b)} the same model with the loop unrolled up to two executions.}
	\label{fig:l1l_example}
\end{figure}
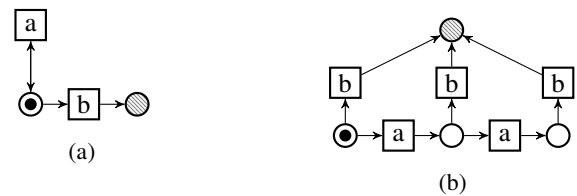
Projected Conformance Checking (PCC) precision was developed by Leemans et al. \cite{Leemans2016} as a computationally efficient precision measure that scales to event logs with billions of events. PCC precision projects both event log and model on all subsets of activities of size $k$, and generates minimal deterministic finite automata (DFA) for the behavior over these subsets of activities in the log (i.e., \emph{log automaton}) and for the behavior over these events allowed by the model (i.e., \emph{model automaton}). Based on the log automaton and model automaton it then builds a \emph{conjunction automaton} which allows the behavior that was allowed both in log and model automaton. It then iterates over the states of the model automaton, and calculates over the share of transitions of this state that is also possible in the corresponding state in the conjunction automaton. It defines precision as the average of this share over the model automaton states.\looseness=-1

PCC precision is implemented in the ProM package ProjectedRecallAndPrecision\footnote{https://svn.win.tue.nl/trac/prom/browser/\\Packages/ProjectedRecallAndPrecision/}. PCC precision assumes the Petri net to be of the class of bounded Petri nets, i.e., the Petri net for which precision is calculated must have a finite number of tokens in every place for all reachable markings.

Consider log $L=[\langle a,b\rangle]$ and Petri nets $M_1$ and $M_2$ of Figures \ref{sfig:l1l} and \ref{sfig:l1l_unrolled} respectively. $M_1$ starts with a length-one-loop on activity $a$, followed by activity $b$. $M_2$ unrolls the length-one-loop on activity $a$ of $M_1$ to at most two executions, thereby limiting the behavior as it only allows at most two executions of activity $a$. It is easy to see that $M_1$ and $M_2$ both belong to the class of bounded Petri nets, as in both models each place can have at most one token. For this log and these models, PCC precision results in $\mathit{prec}(L,M_1){=}0.6$, and $\mathit{prec}(L,M_2){=}0.5$. However, since $\Lan(M_2){\subset}\Lan(M_1)$, \textbf{A2} states that the precision of $M_2$ for fitting log $L$ should be higher or equal to its precision of $M_1$.  This shows that \textbf{A2} does not hold for PCC precision.

This drop in precision is an effect of the additional states that are created in the model DFA as an effect of unrolling the length-one-loop. The model DFA created from Petri net $M_2$ (Figure \ref{sfig:l1l_unrolled}) for example contains a state $s$ that is reached after firing $\langle a,a \rangle$. This state however is never reached based on event log $L$, which only contains a trace $\langle a,b \rangle$, which has the effect that none of the enabled transitions from state $s$ were observed in the log, bringing down the precision. In the DFA generated from Petri net $M_1$ (Figure \ref{sfig:l1l}), this state $s$ is merged with the state that one reaches after observing a single $a$ event, as future behavior allowed by the model does not depend on the number of $a$-events seen.

\begin{figure}
	\centering
	\begin{tikzpicture}
	[node distance=0.7cm,
		on grid,>=stealth',
		bend angle=20,
		auto,
		every place/.style= {minimum size=4mm},
		every transition/.style = {minimum size = 4mm}
	]
	\node [place, tokens = 1, pattern=custom north west lines,hatchspread=1.5pt,hatchthickness=0.25pt,hatchcolor=gray] at (0,0) (p1){};
	\node [transition] (t1) [align=center] at (0.7,0.0) {b} 
	edge [pre] node[auto] {} (p1)
	edge[post] node[auto] {} (p1);
	\node [transition] (t2) [align=center] at (0.0,-0.7) {c} 
	edge [pre] node[auto] {} (p1)
	edge[post] node[auto] {} (p1);
	\node [transition] (t3) [align=center] at (0,0.7) {a} 
	edge [pre] node[auto] {} (p1)
	edge[post] node[auto] {} (p1);
	\end{tikzpicture}
	\caption{A flower model over activities $a$, $b$ and $c$.}
	\label{fig:flower}
\end{figure}

Consider Petri net $M$ of Figure \ref{fig:flower}, and event logs $L_1{=}[\langle b,a,c\rangle,\langle a,a,c\rangle]$, and $L_2{=}[\langle b,a,c\rangle,\langle a,a,c\rangle,\langle a,b,b,b,b,b,b,b,b,b,b,b,b,b,\allowbreak b,b\rangle,\langle b,a,a,a,a,a,a,a,a,a,a,a,a,a,a,a\rangle]$. The single place of $M$ is bounded to one token, therefore $M$ belongs to the class of bounded Petri nets. It is easy to see that $\tilde{L}_1{\subset}\tilde{L}_2$, since the first two traces of log $L_2$ form log $L_1$. PCC precision results in $\mathit{prec}(L_1,M){=}0.3125$ and $\mathit{prec}(L_2,M){=}0.2727$, violating \textbf{A5}. The two traces of $L_2$ that are not in $L_1$ are very long traces to the traces that are in $L_1$, leading to additional states in the log automaton and the conjunction automaton. The additional states of the conjunction automaton have a low precision of $\frac{1}{4}$, since for each state the model allows for four options (firing activity $a$, $b$, $c$, or stopping), while only one is seen in the log. Therefore, if we would expand trace $\langle b,a,a,a,a,a,a,a,a,a,a,a,a,a,a,a\rangle$ with more events of activity $a$, then $\mathit{prec}(L_2,M)$ would approach $\frac{1}{4}$.

\subsection{Overview of Precision Metric Properties}
\begin{table}
	\centering
	\caption{Overview of the precision axioms and whether they hold for each precision measure.}
	\label{tab:overview}
	\scalebox{0.77}{
		\begin{tabular}{|l|c|c|c|c|c|}
			\toprule
			Metric & A1 & A2 & A3 & A4 & A5 \\
			\midrule
			Simple behavioral appropriateness & \xmark & & & \xmark &  \\
			Advanced behavioral appropriateness & \xmark & & \xmark & \cmark & \\
			%Escaping Edges (ETC)     & \xmark & \xmark &  & &&&\\
			One-align ETC            & \xmark & \xmark && \xmark & \xmark \\
			Negative Event Precision & \xmark & \xmark & & &  \\
			PCC precision            &  & \xmark &&  & \xmark \\
			\bottomrule
	\end{tabular}}
\end{table}

We formulated five axioms that describe desirable properties for precision measures. Table \ref{tab:overview} gives an overview of that axioms that we showed that do hold (\cmark) and that do not hold (\xmark) for each precision measure. We found that none of the existing precision measures fulfills all five axioms. Empty cells in the table are currently unknown, and no formal proof nor a counter example has been found that proves or disproves the axiom for the respective precision measure.

\section{Contexts With Unclear Requirements for Precision Metrics}
\label{sec:undefined_situations}
The axioms introduced in Section \ref{sec:axioms} can be regarded as necessary conditions for precision measures, but they leave precision unspecified in some contexts. Figure \ref{sfig:different_models} shows a situation in which $\tilde{L}{\subseteq}\Lan(M_1)$, $\tilde{L}{\subseteq}\Lan(M_2)$, but $\Lan(M_1){\setminus}\Lan(M_2){\ne}\emptyset$ and $\Lan(M_2){\setminus}\Lan(M_1){\ne}\emptyset$. In this setting, both $M_1$ and $M_2$ allow for (a possibly infinite amount of) different behavior that was not seen in $L$. Precision measures deal with this situation by quantifying the amount of behavior of $M_1$ and $M_2$. However, there are no obvious formal properties telling how the precision of $M_1$ and $M_2$ on $L$ should relate.

Furthermore, all axioms define desired properties of precision measures when the event log $L$ fits the behavior of the model $M$, i.e., $\tilde{L}\subseteq\Lan(M)$. In practice, process discovery techniques will return process models with fitness below $1$, i.e., there exists $\sigma{\in} L:\sigma{\notin}\Lan(M)$. The discovery algorithm may deliberately abstract from infrequent behavior. In this paper we do not formulate axioms for precision measures in the context of event logs that do not fit the process model, since we feel that there is not enough agreement in the process mining community on how a precision measure should behave in this context. Figure \ref{sfig:non-fitting_models} shows an Euler diagram of a log $L$ and two models $M_1$ and $M_2$ such that $\Lan(M_1){\subset}\Lan(M_2)$ and $\tilde{L}{\nsubseteq}\Lan(M_1)$, which is a non-fitting equivalent of \textbf{A2}. \textbf{A2} prescribes $\mathit{prec}(L,M_1){\ge}\mathit{prec}(L,M_2)$, however, when the log does not fit the models, the behavior that fits $M_2$ but not $M_1$, $(\tilde{L}{\setminus}\Lan(M_1)){\cap}\Lan(M_2)$, makes it unclear how the precision of $M_1$ and $M_2$ should relate. Furthermore, even when $(\tilde{L}{\setminus}\Lan(M_1)){\cap}\Lan(M_2){=}\emptyset$, it can be the case that the behavior in $L$ that does not fit the models is behaviorally similar to behavior of $M_2$.

\begin{figure}
	\centering
	\begin{subfigure}{0.49\linewidth}
	\vspace{0.35cm}
	\scalebox{0.7}{
		\begin{tikzpicture}
		\node[draw,circle,minimum size=1cm, inner sep=0pt,anchor=west,label={[yshift=0.75cm]below:$\tilde{L}$}] at (1.25,0) {};
		\node[draw,ellipse,minimum height=1.8cm, minimum width=3.5cm,anchor=west,label={[xshift=0.3cm]right:$\Lan(\mathit{M_2})$}] at (-1,0) {};
		\node[draw,ellipse,minimum height=1.8cm, minimum width=3.5cm,anchor=west,label={[xshift=-5cm]right:$\Lan(\mathit{M_1})$}] at (1,0) {};
		\end{tikzpicture}}
		\caption{}
		\label{sfig:different_models}
	\end{subfigure}
	\begin{subfigure}{0.49\linewidth}
	\scalebox{0.7}{
		\begin{tikzpicture}
		\node[draw,ellipse,minimum height=1.5cm, minimum width=1cm,inner sep=0pt,anchor=west,label={[yshift=0.75cm]below:$\tilde{L}$}] at (1,0.5) {};
		\node[draw,ellipse,minimum height=1.6cm, minimum width=3.1cm,anchor=west,label={[xshift=-1.5cm]right:$\Lan(\mathit{M_1})$}] at (0.5,0) {};
		\node[draw,ellipse,minimum height=1.8cm, minimum width=5.2cm,anchor=west,label={[xshift=-1.5cm]right:$\Lan(\mathit{M_2})$}] at (0,0) {};
		\end{tikzpicture}}
		\caption{}
		\label{sfig:non-fitting_models}
	\end{subfigure}
	\caption{Two situations in which the desired properties of precision measures are unclear: \emph{(a)} two models on which log $L$ fits, with both models allowing for behavior that is not allowed by the other model, and \emph{(b)} two models and a log on which the models do not fit.}
	\label{fig:non-fitting}
	\vspace{-0.35cm}
\end{figure}

\section{Conclusions \& Future Work}
\label{sec:conclusion}
In this paper provides a set of minimal requirements for precision measures through axioms. We validated these axioms for existing measures. Surprisingly, we discovered that none of the existing precision measures fulfills all formulated requirements.

In future work, we would like fill the empty cells of Table \ref{tab:overview} and get a complete overview of the axioms that hold for each precision measure. Furthermore, we would like to use the insights learned from evaluating the axioms on the measures to either repair one of the existing measures or come up with a completely new measure that fulfills all five axioms.

\medskip\noindent\textbf{Reproducibility}. The event logs and process models that are used as part of a counterexample for a combination of an axiom and a precision measure can be found at \cite{Tax2017}.
	
%\section*{References}
\bibliographystyle{elsarticle-num}
\bibliography{thebibliography}
\end{document}